\documentclass[preprint,floats,aps,showpacs]{revtex4}
\usepackage{graphicx}

\begin{document}
\newcommand{\beq}{\begin{equation}}
\newcommand{\eeq}{\end{equation}}
\newcommand{\AVG}[1]{\langle{#1}\rangle}
\newcommand{\vev}[1]{\langle{#1}\rangle}
\newcommand{\CH}{{\cal H}}
\def\OMIT#1{{}}
\def\lqcd{\Lambda_{\rm QCD}}

\preprint{\vbox{ \hbox{UCSD/PTH 01--24} \hbox{hep-ph/0112323} \hbox{} }}

\title{Global duality in heavy flavor hadronic decays}

\author{Benjam\'\i{}n Grinstein}

\affiliation{
  Physics Department, University of California at San Diego, 
  La Jolla, California 92093}

\begin{abstract}%
We show that heavy meson hadronic decay widths satisfy quark-hadron
duality when smeared over the heavy quark mass, $M$, to an accuracy of
order $1/M^2$.
\end{abstract}

\date{December, 2001}

\maketitle


Quark-hadron duality has been part of the lore of strong interactions
for three decades. Bloom and Gilman\cite{Bloom:1971ye,Bloom:1970xb}
(BG) discovered duality in electron-proton inelastic
scattering. There, the cross section is given in terms of two Lorentz
invariant form factors $W_1$ and $W_2$ which are functions of the
invariant mass of the virtual photon, $q^2$, and the energy transfer
to the electron, $\nu$.  Considering the form factors as functions of
the scaling variable $\omega\equiv q^2/2M\nu$, they compared the
scaling regime of large $q^2$ (and large $\nu$) with the region of
fixed, low $q^2$. They determined that, for each form factor, the low
$q^2$ curves oscillate about the scaling curve, that identifiable
nucleon resonances are responsible for these oscillations and that the
amplitude of a resonant oscillation relative to the scaling curve is
independent of $q^2$. Moreover, they introduced sum rules whereby
integrals of the form factors at low and large $q^2$ agree and noticed
that the agreement was quite good even when the integration involved
only a region that spans a few resonances.

Poggio, Quinn and Weinberg\cite{Poggio:1976af} (PQW) applied these
ideas to electron-positron annihilation. While BG compared
experimental curves among themselves, PQW compared the experimental
cross section to a scaling curve calculated in QCD. They noticed that
the weighted average of the cross section $\sigma(s)$,
\beq 
\label{eq:pqwavgdefd}
\AVG{\sigma(s)} = {\Delta\over\pi}
\int_0^\infty ds' {\sigma(s')\over (s'-s)^2+\Delta^2} 
\eeq 
is given in terms of the vacuum polarization of the electromagnetic
current with complex argument,
\beq 
\AVG{\sigma(s)} = {1\over2i}\Big(\Pi(s+i\Delta)-\Pi(s-i\Delta)\Big), 
\eeq 
and argued that one can safely use perturbation theory to compute this
provided $\Delta$ is large enough. This procedure was better
understood with the advent of Wilson's\cite{Wilson:1969} Operator
Product Expansion (OPE).  It is interesting to point out that the
prediction of PQW based on the two generations of quarks and leptons
known at the time did not successfully match the experimental
results. When PQW allowed for additional matter they found a best
match if they supplemented the model with a heavy lepton and a charge
$1/3$ heavy quark, anticipating the discoveries of the tau-lepton and
b-quark. 

In an attempt to understand the origin of quark-hadron duality we have
computed both the actual rate and its ``scaling limit'' from first
principles in special situations.  In Ref.~\cite{Boyd:1996ht} we
computed the semi-leptonic decay rate and spectrum for a heavy hadron
in the small velocity (SV) limit. We showed that two channels, $B\to
De\nu$ and $B\to D^*e\nu$, give the decay rate to first two orders in
an expansion in $1/m_b$ and that to that order the result is identical
to the inclusive rate obtained using a heavy quark OPE as introduced
in Ref.~\cite{Chay:1990da}. The equality holds for the double differential
decay rate if it is averaged over a large enough interval of hadronic
energies. The computation demonstrates explicitly quark-hadron duality
in semi-leptonic $B$-meson decays in the SV limit, but really sheds no
light into the mechanism for duality. In particular, it is puzzling
that duality holds even if the rate is dominated by only two 
channels.

More recently we attempted to verify duality in hadronic heavy meson
decays. In Ref.~\cite{Grinstein:1998xk} we considered the width of a
heavy meson in a soluble model that in many ways mimics the dynamics
of QCD, namely an $SU(N_c)$ gauge theory in $1+1$ dimensions in the
large $N_c$ limit. This model, first studied by
't~Hooft\cite{'tHooft:1974hx}, exhibits a rich spectrum with an
infinite tower of narrow resonances for each internal quantum number,
making the study of duality viable. We considered a `$B$-meson' with a
heavy quark $Q$ and a light (anti-)quark $q$ of masses $M_Q$ and $m$,
respectively, which decays via a weak interaction into light $\bar q
q$ mesons. To leading order in $1/N_c$ the decay rate is dominated by
two body final states: if $\pi_j$ denote the tower of $\bar q
q$-mesons, the total width is given by
$\Gamma(B)=\sum\Gamma(B\to\pi_j\pi_k)$, where the sum extends over all
pairing of mesons such that the sum of their masses does not exceed
the $B$ mass, $\mu_j+\mu_k<M_B$. The main result of that investigation
was that there is rough agreement between $\Gamma(B)$ and the decay
rate of a free heavy quark, $\Gamma(Q)$. When considered as functions
of $M_Q$ the quark rate is smooth but the meson rate exhibits sharp
peaks whenever a threshold for production of a light pair opens
up. This is due to the peculiar behavior of phase space in $1+1$
dimensions, which is inversely proportional to the momentum of the
final state mesons. Nevertheless, in between such peaks it was found
that the relation $\Gamma(B)=\Gamma(Q)(1+0.14/M_Q)$, in units of
$g^2N_c/\pi=1$, holds fairly accurately.

Recently\cite{Grinstein:2001zq} we considered the effect of local
averaging on the results of Ref.~\cite{Grinstein:1998xk}. The main
result is that when averaged locally over the heavy mass $M_Q$ the
agreement between $\Gamma(B)$ and $\Gamma(Q)$ is parametrically
improved. In fact, for the averaged widths we found
\beq
\label{eq:mainresult}
\AVG{\Gamma(B)}\approx\AVG{\Gamma(Q)}\left[1+{0.4\over M_Q^2}+{5.5\over
M_Q^3}\right] 
\eeq 
Remarkably, the correction of order $1/M_Q$ has disappeared.

In this paper we demonstrate that when averaging over $M_Q$ the
corrections of order $1/M_Q$ are absent. The argument we present is
very general and applies both to the 't~Hooft model, explaining the
numerical observations of \cite{Grinstein:2001zq}, and the
phenomenological relevant case of four dimensional QCD. The central
idea is simple. In a heavy quark effective theory the four quark
operator describing a weak $B$-meson hadronic decay, is
\beq
e^{-iMv\cdot x}(\bar u_L\gamma^\mu h_v \bar d_L\gamma_\mu u_L)(x),
\eeq
where $h_v$ is the heavy quark field with velocity $v$. The
exponential factor, which accounts for the large momentum carried by
the heavy quark, plays the same role as an insertion of external
momentum $\exp(-iq\cdot x)$ with the specific choice $q=Mv$. Thus one
can use dispersion relations to relate the decay amplitude to Green
functions with complex momentum where an OPE is valid, much like the
procedure for semileptonic decays in Ref.~\cite{Chay:1990da}. The resulting
relation has then the form of a {\it mass averaged} amplitude in terms
of a systematic OPE. 

\begin{figure}
\includegraphics[scale=.7]{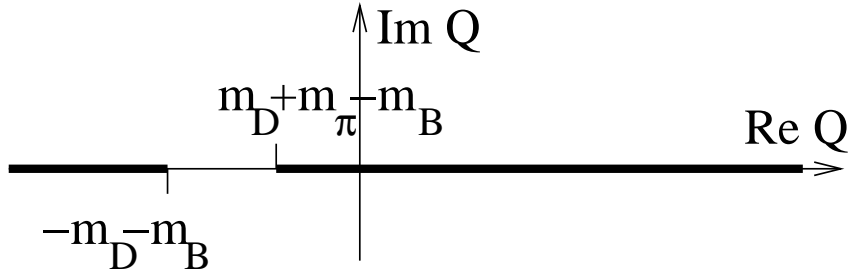}
\caption{Analytic strcuture of the Green function $T(Q)=T(Mv,Qv)$ in
the complex $Q$ plane. Cuts on the real line are depicted by a heavy
solid line.}
\label{fig:fullthcuts}
\end{figure}

Consider the Green function
\beq
T(q,p)=i\int d^4x\;e^{iq\cdot x}
\vev{\bar B(p)|T(\CH^\dagger(x)\CH(0))|\bar B(p)}
\eeq
where the $B$ momentum is $p=Mv$ and $\CH$ is the term in the weak
Hamiltonian density responsible for hadronic $B$ decay:
\beq
\CH={4G_F \over\sqrt2}V_{cb}^{\phantom{\dagger}}V_{ud}^*
\bar c_L\gamma^\mu b \bar d_L\gamma_\mu u_L.
\eeq
A simple calculation gives
\begin{eqnarray}
\mbox{Im}T=\sum_X\pi(2\pi)^3\delta^4(q+p-p_X)|\vev{X|\CH(0)|\bar
B}|^2\\
+\sum_X\pi(2\pi)^3\delta^4(q-p+p_X)|\vev{X|\CH^\dagger(0)|\bar B}|^2
\end{eqnarray}
Hence, the analytic structure of $T(Q)\equiv T(Mv,Qv)$ is as shown in
Fig.~\ref{fig:fullthcuts}. The two real axis cuts are associated with
the two time orderings of $\CH$ and $\CH^\dagger$. The discontinuity
across the first cut, which runs from $Q=Q_1=m_D+m_\pi-m_B$ to
infinity, is related to the inclusive decay rate of the $\bar B$
meson. For $Q>m_B$ there is also a contribution from states with two
$\bar B$ mesons. The discontinuity across the second cut, running from
$Q=Q_2=-m_B-m_D$ to negative infinity, is related to a process with
two units of $B$-number in the final state. In addition, a pole at
$-m_{B_c}$ is not shown. The decay rate is obtained as the
discontinuity at $q=0$,
\beq
\Gamma(B)={1\over m_B}\mbox{Im}T|_{q=0}.
\eeq
While $T$ may be computed perturbatively when the complex momentum
$q^0$ is sufficiently away from the real cut, the  computation of $\Gamma(B)$
requires $T$ at one point on the cut itself. This has been the main
impediment to computing the decay width. In processes such as $e^+e^-$
annihilation into hadrons or in semileptonic $B$ decays, an
integration over $q$ allows one to use a dispersion relation that
relates an integral of the discontinuity of $T$ on the real axis to
the value of $T$ in the complex plane. But in this process $q$ is
fixed.

Our solution to this problem makes use of the observation above that
when computing $T$ in an effective theory for the static heavy quark
the momentum of the heavy quark, $Mv$, and the external insertion of
momentum, $q$, enter all expressions in the precise combination
$q+Mv$. Therefore, one may still use a dispersion relation integrating
over $q$, and this will have the same effect as an integral over
$M$. The result is a perturbative expression for an integral over the
mass of the decay width.

We define a Green function in the effective theory similarly,
\beq
\label{eq:greeneff}
\hat T(q,v)=i\int d^4x\;e^{i(q+Mv)\cdot x}
\vev{H(v)| T(\hat{\CH}^\dagger(x)\hat\CH(0))|H(v)}
\eeq
Here $H(v)$ is the state corresponding to the static quark with four
velocity $v$, with a non-standard normalization (independent of $M$)
as is appropriate in the effective theory. To this order, the weak
Hamiltonian in the effective theory, $\hat \CH$, is  the weak
Hamiltonian $\CH$  with the quark $b$ replaced by the effective
theory static field $h_v$. It follows that
\beq
\mbox{Im}\hat T=\sum_X\pi(2\pi)^3\delta^4(q+Mv-p_X)|\vev{X|\hat\CH(0)|
H}|^2.
\eeq
A second term, of the form
\beq
\sum_X\pi(2\pi)^3\delta^4(q+Mv+p_X)|\vev{X|\hat\CH^\dagger(0)|H}|^2,
\eeq
is absent because the intermediate state $X$ is required to have two
units of $B$-number, which is excluded from the effective theory. This
also means that the left cut in Fig.~\ref{fig:fullthcuts} is absent for
$\hat T$. The effective theory, while unable to properly reproduce the
full Green function $T(p,q)$, does provide a systematic approximation
to physical quantities like the hadronic decay width, thus
$\Gamma(B)\approx \mbox{Im}\hat T|_{q=0}$.

\begin{figure}
\includegraphics[scale=.5]{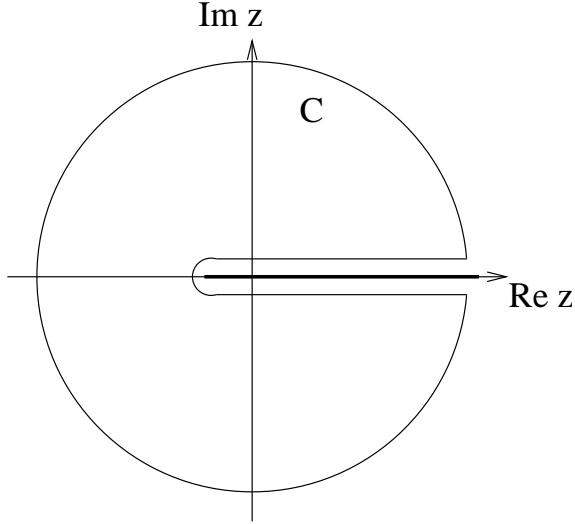}
\caption{Contour for the integral in the complex $Q$ plane leading to
Eq.~(\protect\ref{eq:toff}). The thick line on the real axis denotes
the cut of the Green function $\hat T(Q)$. }
\label{fig:contour}
\end{figure}

To leading order in $1/M$ one has an additional result: since the mass
enters only in the combination $q+Mv$, the decay rate for a heavy
meson of heavy quark mass $M+\delta M$ is $\Gamma(M+\delta M)\approx
\mbox{Im}\hat T|_{q=\delta Mv}$. If we define $\hat T(Q)=T(Qv,v)$,
which depends  implicitly on $M$, then to leading order in $M$ the
dependence on $M$ and $Q$ is only through the combination $M+Q$. 
It is straightforward now to use standard methods of analysis to
relate an integral of $\Gamma$ to the Green function of complex
argument. 
Using the contour in Fig.~\ref{fig:contour}, we have
\beq
\label{eq:toff}
\left.{1\over(n-1)!}{d^{n-1}\over dz^{n-1}}
 {\hat T(z)\over(z+i\Delta)^n}\right|_{z=i\Delta}+
\left.{1\over(n-1)!}{d^{n-1}\over dz^{n-1}}
 {\hat T(z)\over(z-i\Delta)^n}\right|_{z=-i\Delta}=
{1\over2\pi i}\oint dz  {\hat T(z)\over(z^2+\Delta^2)^n}.
\eeq
The right hand side of this equation is the width calculated to
leading order in $1/M$ in the effective theory, $\hat \Gamma_0$,
averaged over masses with a particular weight. We have introduced a
parameter $n$, the power of the denominator in (\ref{eq:toff}), to
guarantee vanishing of the integral on the circle at infinity. It
needs to be adjusted depending on the number of spacetime
dimensions. Defining
\beq
\label{eq:AVGdefd}
\AVG{f(M)}=\int_{-\infty}^\infty dx\;w(x)f(x)
\eeq
with the weight function defined by
\beq
\label{eq:wdefd}
w(x)={(n-1)!\over(2n-3)!!}{1\over 2\pi\Delta}
\left({2\Delta^2\over(x-M)^2+\Delta^2}\right)^n,
\eeq
and recalling that the width vanishes when $m_B<m_D+m_\pi$, we have obtained
\beq
\label{eq:AVGgamma}
\AVG{\hat\Gamma_0(M)}={2^{n-1}\Delta^{2n-1}\over(2n-3)!!}\left\{\left.{d^{n-1}\over
dz^{n-1}}{\hat T(z)\over(z+i\Delta)^n}\right|_{z=i\Delta} +
\left.{d^{n-1}\over
dz^{n-1}}{\hat T(z)\over(z-i\Delta)^n}\right|_{z=-i\Delta}\right\}.
\eeq
This is of the form of our main result, but is not quite complete. It
gives a peculiar average over heavy quark mass of the decay width to
leading order in $1/M$ in terms of the off-shell effective theory
Green function with complex momentum.  The right hand side can be
evaluated using an operator product expansion if $\Delta$ is large
enough. As in the case of semileptonic decays\cite{Chay:1990da} the
leading operator in the expansion is of the form $\bar h_v\Gamma h_v$
and has a normalized matrix element in the state $H(v)$. Since its
coefficient is computed perturbatively, one has, to leading order in
$1/M$,
\beq
\label{eq:AVGequalAVG}
\AVG{\hat\Gamma_0(M)}=\AVG{\Gamma_Q(M)}
\eeq
where $\Gamma_Q(M)$ is the perturbative width of the heavy quark.

The next term in the operator product expansion of $\hat T(i\Delta)$
involves an operator with one derivative, $\bar h_v \Gamma D_\mu h_v$,
which has vanishing expectation value in the state $H(v)$. Therefore,
the leading correction to Eq.~(\ref{eq:AVGequalAVG}) from the OPE is
of order $1/M^2$. However, the Hamiltonian $\CH$ itself has an
expansion in $1/M$ so we have yet to establish the validity of
quark-hadron duality for the decay width to order $1/M^2$. Moreover,
the relation between the weak Hamiltonian and its representation in
the HQET involves Wilson coefficients that have explicit mass
dependence. Both types of correction spoil the invariance $M\to
M+\delta M$, $Q\to Q-\delta M$. We address these issues next.

\begin{figure}
\includegraphics[scale=0.5]{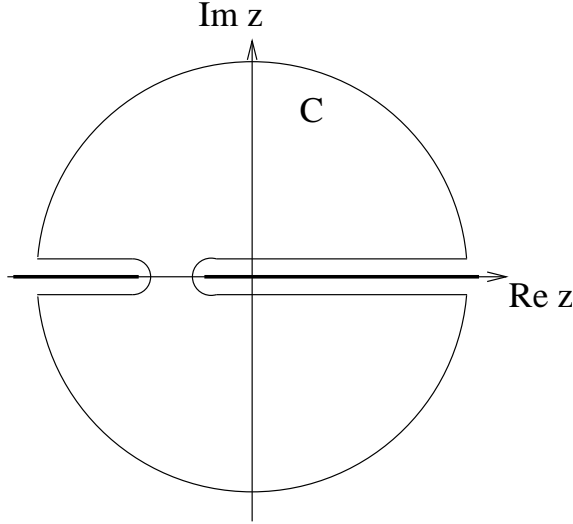}
\caption{Contour for the integral in the complex $Q$ plane leading to
Eq.~(\protect\ref{eq:toffm}). The thick lines on the real axis denote
the cuts in the Green function $\hat T_k$ and coefficient function $C_k$. }
\label{fig:contour2}
\end{figure}

The effective Hamiltonian $\CH$ has an HQET expansion in powers of
$1/M$,
\beq
\label{eq:hatHexpand}
 \CH = C_0\hat\CH_0 +{1\over M}C_1\hat\CH_1+\cdots
\eeq
A sum over several possible operators at each order in $1/M$ is implicit. 
The coefficients $C_k$ are functions of $M/\mu$, where $\mu$ is a
renormalization point which we chose to be a fixed number, large enough that
the coefficients can be computed perturbatively. We do not set $\mu=M$ since
this would introduce additional $M$ dependence into the operators
(which are also renormalized at the scale $\mu$). 
There is a corresponding expansion of the Green function in
Eq.~(\ref{eq:greeneff}), 
\beq
T=C_0\hat T_0 + {1\over M}C_1\hat T_1 +\cdots
\eeq
and of the width,
\beq
\Gamma = \hat \Gamma_0 + \hat \Gamma_1 + \cdots
\eeq
Consider the individual averages
\beq
\AVG{\hat\Gamma_k(M)}=\int dQ\; w(Q)\hat \Gamma_k(Q)
\eeq
where the weight function $w$ is given in Eq.~(\ref{eq:wdefd}). In
order to use a dispersion relation like in~(\ref{eq:toff}) we note that the
explicit inverse powers of mass give poles at $z=-M$ and the Wilson
coefficients, with typical $\ln(M)$ behavior,  give cuts extending
from $z=-M$ to $-\infty$. Using the contour  in
Fig.~\ref{fig:contour2}, which  excludes these cut
and pole, we are led to consider
\begin{eqnarray}
\label{eq:toffm}
& &{1\over 2\pi i }\oint dz\; {C_k(z+M)\hat T_k(z) \over
(z+M)^k(z^2+\Delta^2)^n}=\\
& &~~\left.{1\over(n-1)!}{d^{n-1}\over dz^{n-1}}
 {C_k(z+M)\hat T_k(z)\over(z+M)^k(z+i\Delta)^n}\right|_{z=i\Delta}+
\left.{1\over(n-1)!}{d^{n-1}\over dz^{n-1}}
 {C_k(z+M)\hat T_k(z)\over(z+M)^k(z-i\Delta)^n}\right|_{z=-i\Delta}\nonumber
\end{eqnarray}
On the other hand, the integral can be written as a sum of two terms,
namely, the width average we want and the integral over a contour
below and above the cut on the negative real axis. The latter is
suppressed by powers of $M$. To estimate it we note that since the
Green function $\hat T$ is analytic in this region we may simply
replace it by a power of the mass given by dimensional analysis, $\hat
T(z)\sim (z+M)^p$ where $p=5$ in four dimensions ($p=2D-3$ in the
general case of $D$ dimensions). Also, we may take the Wilson
coefficient to be a simple $\log$ for this estimate. Then the integral
around the cut is
\begin{eqnarray}
{1\over2\pi i }\int &dz&{\ln(z+M)(z+M)^{p-k}\over(z^2+\Delta^2)^n}\nonumber\\
&=&
{1\over(n-1)!}{d^{n-1}\over
dz^{n-1}}\left[\left.{\ln(z+M)(z+M)^{p-k}\over(z+i\Delta)^n}\right|_{z=i\Delta}+(\Delta\to-\Delta)\right]\nonumber\\
&=&{\cal O}(M^{p-k+1-2n}\ln(M))
\end{eqnarray}
For comparison, a similar estimate of the average $\AVG{\Gamma_k(M)}$
using $\Gamma_k(x)\sim x^{p-k}$ gives
\beq
\int_0^{\infty}dx{x^{p-k}\over[(x-M)^2+\Delta^2]^n}
={2\pi\over(2\Delta)^{2n-1}} {(2n-2)!\over[(n-1)!]^2}M^{p-k}\left(1+{\cal
O} ({\Delta\over M})\right)
\eeq
Thus, we can express the width average in terms of the off-shell Green
function and derivatives in Eq.~(\ref{eq:toffm}) up to corrections
suppressed by $(\Delta/M)^{2n-1}\ln(M)$. 

As in the case of semileptonic decays\cite{Chay:1990da}, we can now
calculate the width average $\AVG{\Gamma(M)}$ by computing the
off-shell Green functions using an OPE. But we can do better: we can
show that there are no corrections of order $1/M$. As in the case of
semileptonic decays, the OPE is an expansion in operators of the form
$\bar h_v\Gamma D^l h_v$ where $D$ is a covariant derivative. The
central observations are that the operator $\bar\Gamma h_v D_\mu h_v$
has vanishing expectation value\cite{Chay:1990da} and the expansion
for $\hat T_k$ starts at order $\bar h_v \Gamma D^k h_v$. The latter
statement is non-trivial. Consider the case of semileptonic decays. In
the original work of Ref.~\cite{Chay:1990da} this question is
sidestepped by doing a simultaneous expansion in large momentum
transfer (the OPE) and the HQET. But, following
Ref.~\cite{Mannel:1993su}, one could first express the Green function
in the HQET and only then do the OPE. Although
Ref.~\cite{Mannel:1993su} does not consider the effect of subleading
operators, it is clear that the two approaches yield the same result
only if the OPE of products of subleading operators starts at the
corresponding order in $\bar h_v D^l h_v$. Clearly, a similar indirect
argument applies here, but we know of no direct proof of the
statement.

Our main result is then
\beq
\AVG{\Gamma(M)}=\AVG{\Gamma_Q(M)}\left(1+{\cal O}({1\over M^2})\right).
\eeq
It should be noted that the corrections that have been omitted are
parametrically small at large $M$, but can be quantitatively large,
depending on the values of $M$, $n$ and $\Delta$. 

In the 't~Hooft model our result is in agreement with the empirical
observations of Ref.~\cite{Grinstein:2001zq} where the averaged widths
agree to order $1/M^2$ while the un-averaged ones agree at best to
order $1/M$. The weight functions used for averages in
Ref.~\cite{Grinstein:2001zq} were Gaussian, $w(x)\sim
x^n\exp(-(x-M)^2/\Delta^2)$. We have checked that the results still
hold for the weight function in Eq.~(\ref{eq:wdefd}). It is
interesting that substantial duality violation is found if the power
$n=1$ is used. In the realistic case of QCD in four dimensions one is
left with the very realistic possibility that the physical hadronic
width of a heavy meson exhibits oscillations of magnitude $1/M$ about
the partonic width which are erased out when performing unphysical
mass averages. Some evidence for this was presented in
Ref.~\cite{Altarelli:1996gt} where it was observed that the $b$-quark
width agrees better with experimental hadronic widths if the quark
mass is replaced by the $B$ or $\Lambda_b$ masses, respectively, and
in Ref.~\cite{Nussinov:2001zc} which argues that the $D^0-D_s$
lifetime difference is also primarily a phase space effect. In a
similar vein, Ref.~\cite{Colangelo:1997ni} shows how $1/M$ violations
to local, but not global, duality may occur in $B$-meson correlations.

To summarize, we have shown that the hadronic width of a heavy meson
averaged over the heavy quark mass as in Eq.~(\ref{eq:AVGdefd}) is
correctly given by the corresponding average of a perturbative heavy
quark width up to corrections of order $1/M^2$. This result can be
applied to the decay widths of heavy mesons in the 't~Hooft model, and
explains the numerical observations of
Ref.~\cite{Grinstein:2001zq}. The result, however, is not of direct
phenomenological significance since it is impossible to perform mass
averages of observed decay widths of $B$ mesons. However, our result
adds to the body of evidence that heavy meson widths cannot be
reliably computed using perturbation theory, at least not with a
precision of order $1/M^2$.

\acknowledgments{
This work is supported in part by the Department of Energy
under contract No.\ DOE-FG03-97ER40546.
}

\end{document}